\documentclass[a4paper,11pt]{article}
\usepackage{pos}

\usepackage{slashed}

\hypersetup{
    pdftitle = {Localised Dirac eigenmodes and Goldstone's
      theorem at finite temperature},
    pdfauthor = {Matteo Giordano}
}

\newcommand{\f}[2]{\frac{#1}{#2}}
\newcommand{\tf}[2]{{\textstyle \frac{#1}{#2}}}

\newcommand{\de}{\partial}
\newcommand{\la}{\langle}
\newcommand{\ra}{\rangle}
\newcommand{\lla}{\la\!\la}
\newcommand{\rra}{\ra\!\ra}
\newcommand{\Oc}{O}
\newcommand{\Ac}{{A}}
\newcommand{\Pc}{{P}}

\newcommand{\Gc}{{\cal G}}

\newcommand{\tr}{{\rm tr}\,}
\newcommand{\chicond}{\Sigma}

\title{Localised Dirac eigenmodes and Goldstone's theorem at finite
  temperature}

\author*[a]{Matteo Giordano}

\affiliation[a]{ELTE E\"otv\"os Lor\'and University, Institute for
  Theoretical Physics,\\ P\'azm\'any P\'eter s\'et\'any 1/A, H-1117, Budapest,
  Hungary}

\emailAdd{giordano@bodri.elte.hu}

\abstract{I show that a finite density of near-zero localised Dirac
  modes can lead to the disappearance of the massless excitations
  predicted by the finite-temperature version of Goldstone's theorem
  in the chirally broken phase of a gauge theory.}

\FullConference{%
 The 38th International Symposium on Lattice Field Theory, LATTICE2021
  26th-30th July, 2021
  Zoom/Gather@Massachusetts Institute of Technology
}

\begin{document}
\maketitle

\section{Introduction}
\label{sec:intro}

Ample evidence from lattice calculations shows that the lowest modes
of the Euclidean Dirac operator $\slashed{D}$ are localised in the
high-temperature phase of
QCD~\cite{GarciaGarcia:2006gr,Kovacs:2012zq,Cossu:2016scb,Holicki:2018sms}
and of other gauge theories~\cite{Kovacs:2009zj,Kovacs:2010wx,
  Giordano:2016nuu,Kovacs:2017uiz,Giordano:2019pvc,Vig:2020pgq,
  Bonati:2020lal,Baranka:2021san,Cardinali:2021fpu}
(see~\cite{Giordano:2021qav} for a recent review). Localised modes are
supported essentially only in a finite spatial region whose size does
not change as the system size grows. In contrast, delocalised modes
extend over the whole system and keep spreading out as the system size
is increased. The distinction is made quantitative by the scaling with
the spatial volume $V$ of the \textit{inverse participation ratio}
(IPR), which for a normalised eigenmode $\psi_n(x)$ reads
\begin{equation}
  \label{eq:IPR}
  {\rm IPR}_n = 
  \int_T d^{{\rm d}+1}x\,  \Vert \psi_n(x)\Vert^4\,, \qquad\text{with}\quad
  \int_T d^{{\rm d}+1}x\,  \Vert \psi_n(x)\Vert^2=1\,, 
\end{equation}
where ${\rm d}$ is the spatial dimension of the system,
$\Vert \psi_n(x)\Vert^2=\sum_{A,c} |\psi_{n\,A,c}(x)|^2$ is the local
amplitude squared of the mode summed over colour ($c$) and Dirac ($A$)
indices, and
$ \int_T d^{{\rm d}+1}x = \int_0^{\f{1}{T}} dt \int d^{\rm d}x $, with
$T$ the temperature of the system.  Assuming that $\psi_n(x)$ is
non-negligible only in a region of size $\Oc(V^\alpha)$, one can
easily estimate that ${\rm IPR}_n \sim V^{-\alpha}$. For localised
modes $\alpha=0$, while for delocalised modes $0<\alpha\le 1$.

Lattice studies show the same situation in a variety of gauge
theories, with different gauge groups and in different dimensions
(also using different fermion discretisations): while delocalised in
the low-temperature, confined phase, low modes are localised in the
high-temperature, deconfined phase up to some critical point
$\lambda_c$ in the spectrum, above which they are again delocalised.
Localisation is a well-known phenomenon in condensed matter physics,
commonly appearing in disordered
systems~\cite{lee1985disordered}. Technically, the Dirac operator can
indeed be seen as ($i$ times) the Hamiltonian of a disordered system,
with disorder provided by the fluctuations of the gauge fields.  It is
then not surprising that the features of localisation observed in
gauge theory are analogous to those found in condensed matter systems:
for example, at the ``mobility edge'' $\lambda_c$, where localised
modes turn into delocalised modes, one finds a second-order phase
transition along the spectrum (``Anderson
transition'')~\cite{Giordano:2013taa} with critical spectral
statistics~\cite{Nishigaki:2013uya} and multifractal
eigenmodes~\cite{Ujfalusi:2015nha}, exactly as in condensed matter
systems~\cite{Evers:2008zz}.

The physical consequences of localisation in disordered systems are
clear: most notably, localisation of electron eigenmodes leads to the
transition from conductor to insulator in a metal with a large amount
of impurities~\cite{lee1985disordered}. The situation is instead not
so clear for gauge theories, where the physical meaning of the
localisation of Dirac modes has proved to be more elusive. There is,
however, growing evidence of an intimate connection between
localisation and deconfinement: in a variety of systems with a genuine
deconfinement transition, localisation of the low Dirac modes appears
in fact precisely at the critical
point~\cite{Giordano:2016nuu,Kovacs:2017uiz,Giordano:2019pvc,Vig:2020pgq,
  Bonati:2020lal,Baranka:2021san,Cardinali:2021fpu}. This is true even
for the simplest model displaying a deconfinement transition, namely
2+1 dimensional $\mathbb{Z}_2$ gauge theory~\cite{Baranka:2021san}.
Theoretical arguments for this behaviour have also been discussed in
the
literature~\cite{Bruckmann:2011cc,Giordano:2015vla,Giordano:2016cjs}.
This connection could help in better understanding confinement and the
deconfinement transition.

Still, one would like to find a more direct physical interpretation
for localisation in gauge theories. This may seem a hopeless task,
given that no physical meaning is attached to individual points, or
even regions, of the Dirac spectrum, with observables obtained only
integrating over the whole spectrum.  A notable exception to this
state of affairs is the chiral limit: in this case the point
$\lambda=0$ is singled out as only near-zero modes are physically
relevant, and the localisation properties of these modes may have
direct physical implications.  In particular, one wonders how a finite
density of near-zero localised modes can affect (if at all) the usual
picture of spontaneous chiral symmetry breaking and generation of
Goldstone excitations.  While I know of no model where such a scenario
has been demonstrated, there are intriguing hints in (i) 2+1 flavour
QCD towards the chiral limit, and (ii) SU(3) gauge theory with $N_f=2$
massless adjoint fermions.
\begin{itemize}
\item[(i)] A peak of localised near-zero modes has been observed in
  overlap spectra computed in HISQ backgrounds for near-physical
  light-quark mass right above the crossover temperature
  $T_c$~\cite{Dick:2015twa}. This peak persists also for
  lighter-than-physical light-quark
  masses~\cite{Ding:2020xlj,Kaczmarek:2021ser}, but the localisation
  properties are not known in that case.  It is possible that this
  peak will survive and the localised nature of the modes will not
  change in the chiral limit.
\item[(ii)] SU(3) gauge theory with $N_f=2$ massless adjoint fermions
  displays an intermediate, chirally broken but deconfined
  phase~\cite{Karsch:1998qj,Engels:2005te}, where a nonzero density of
  near-zero Dirac modes is certainly present. As the theory is
  deconfined, one expects these modes to be localised.
\end{itemize}

\section{Localised modes and Goldstone's theorem at zero temperature}
\label{sec:LGT_T0}

It is instructive to discuss first the case $T=0$. Consider a gauge
theory with $N_f$ degenerate flavours of fundamental quarks of mass
$m$.  In such a theory, as a consequence of the Banks-Casher
relation~\cite{Banks:1979yr} and of Goldstone's
theorem~\cite{Goldstone:1962es}, a nonzero density of near-zero modes
in the chiral limit implies the spontaneous breaking of chiral
symmetry down to SU$(N_f)_V$, and in turn the presence of massless
pseudoscalar Goldstone bosons in the particle spectrum.  However, one
should say more precisely ``delocalised near-zero modes'': in fact, it
has been known for quite some
time~\cite{McKane:1980fs,Golterman:2003qe} that if the near-zero modes
are localised then the Goldstone bosons disappear.  To see this in the
case at hand, one uses the SU$(N_f)_A$ (axial nonsinglet)
Ward-Takahashi (WT) identity,
\begin{equation}
  \label{eq:WI1}
  -\la\de_\mu \Ac^a_\mu(x) \Pc^b(0)\ra + 2m \la \Pc^a(x)\Pc^b(0)\ra =
  \delta^{(4)}(x)\delta^{ab}  \Sigma
  \,,
\end{equation}
where $\Ac^a_\mu =\bar{\psi}\gamma_\mu\gamma_5t^a\psi$,
$\Pc^a_\mu =\bar{\psi}\gamma_5t^a\psi$, and
$\Sigma = \tf{1}{N_f}\la \bar{\psi}\psi\ra$, with $\gamma_\mu$ and
$\gamma_5$ the Euclidean Hermitian gamma matrices and $t^a$ the
generators of SU$(N_f)$ in the fundamental representation normalised
as $2\,\tr t^a t^b = \delta^{ab}$, and $\la\ldots\ra$ is the Euclidean
expectation value. In momentum space Eq.~\eqref{eq:WI1} becomes
\begin{equation}
  \label{eq:WI2}
   ip_\mu \Gc_{AP\mu}(p) + 2m \Gc_{PP}(p) = \Sigma\,,
\end{equation}
where
$\delta^{ab}\Gc_{PP}(p) = \int d^4x\, e^{ip\cdot x} \la
\Pc^a(x)\Pc^b(0)\ra $, and similarly for $\Gc_{AP\mu}(p)$. 
In the limit $m\to 0$, one finds near $p=0$ that
\begin{equation}
  \label{eq:WI3}
  \Gc_{AP\mu}(p)  \mathop\to_{p\to 0} 
  -\f{ip_\mu }{p^2}[\chicond  -{\rm R}]\,,  \qquad
   {\rm R} \equiv \lim_{p\to 0}\lim_{m\to 0} 2m\Gc_{PP}(p) \,,  
\end{equation}
with $\chicond$ denoting from now on the chiral condensate in the
chiral limit. If $\chicond -{\rm R} \neq 0$, $\Gc_{AP\mu}$ has a pole
at zero momentum implying the existence of massless bosons. If
$\Gc_{PP}$ behaves reasonably as a function of $m$ in the chiral limit
then ${\rm R} =0$, and massless bosons are present if chiral symmetry
is spontaneously broken by a nonzero chiral condensate $\chicond$.
However, as I show below in Section~\ref{sec:locpp}, if there is a
finite density of localised near-zero modes then $\Gc_{PP}$ generally
diverges like $1/m$ in the chiral limit.  This divergence leads to a
nonzero ${\rm R}$ proportional to the density of localised near-zero
modes, by cancelling the factor of $m$ in a way reminiscent of how UV
anomalies are formed.  In particular, if a finite mobility edge is
found in the chiral limit then the ``anomalous remnant'' ${\rm R}$
cancels $\chicond$ exactly, removing the pole from $\Gc_{AP\mu}$, and
so the Goldstone bosons from the spectrum.

A non-vanishing anomalous remnant allows one to evade Goldstone's
theorem. In fact, the anomalous remnant leads to chiral symmetry being
{\it explicitly} broken in the chiral limit, with the resulting
modification of the usual WT identity showing that the axial-vector
current is not conserved. Current conservation is a fundamental
hypothesis of the theorem, and since it does not hold the theorem does
not apply.

\section{Localised modes and Goldstone's theorem at finite
  temperature}
\label{sec:LGT_T}

The argument discussed above is not really relevant to realistic gauge
theories (e.g., QCD and QCD-like theories), where no localised
near-zero modes have been observed at $T=0$. However, it suggests a
general strategy to study the physical effects of localisation in the
chiral limit also at finite temperature: relate the properties of the
Euclidean Dirac spectrum with those of the physical spectrum using the
axial nonsinglet WT identity Eq.~\eqref{eq:WI1}, that holds also at
$T\neq 0$.  In this case, due to technical reasons related to the
breaking of O(4) invariance in the Euclidean setting, the physical
spectrum is accessed more naturally by reconstructing the
axial-vector-pseudoscalar spectral function $\varrho^{\Ac\Pc}$ (see
\cite{Meyer:2011gj}) from the Euclidean correlators,
\begin{equation}
  \label{eq:spec_func}
  \varrho^{\Ac\Pc}(\omega,\vec{p}\hspace{0.025em}) \equiv 
  \int d^4x\,
  e^{i(\omega t - \vec{p}\cdot\vec{x})} \lla [\hat{\Ac}^a_0(t,\vec{x}),\hat{\Pc}^b(0)]\rra_T\,,
\end{equation}
where $\lla \ldots\rra_T$ denotes the (real time) thermal expectation
value, and $\hat{\Ac}^a_\mu$ and $\hat{\Pc}^b$ are the Minkowskian
axial-vector and pseudoscalar operators.  Using the WT identity
Eq.~\eqref{eq:WI1} and the symmetry and analyticity properties of the
correlation functions,\footnote{It is also assumed that there is no
  transport peak in the pseudoscalar channel. This is expected on
  general grounds, and supported by numerical lattice results
  (see~\cite{Burnier:2017bod}).}  one finds in the chiral limit at
zero momentum~\cite{Giordano:2020twm}
\begin{equation}
  \label{eq:spec_func2}
  \lim_{\vec{p}\to 0} \lim_{m\to 0} \varrho^{\Ac\Pc}(\omega,\vec{p}) =
  -2\pi  [ \chicond - {\rm R} ] \delta(\omega) + 
  \text{(regular at $\omega=0$)}\,, 
\end{equation}
where $ \chicond$ and ${\rm R}$ are now computed at finite
temperature, i.e., compactifying the Euclidean time direction to size
$1/T$, and in particular
\begin{equation}
  \label{eq:remnant_finiteT}
 {\rm R} = \lim_{\vec{p}\to 0}\lim_{m\to 0}
2m\Gc_{PP}(\omega=0,\vec{p}\hspace{0.025em}) \,.
\end{equation}
The Dirac delta in Eq.~\eqref{eq:spec_func2} indicates the presence of
massless \textit{quasi-particle} excitations in the spectrum, as long as
its coefficient is nonzero.  Similarly to the zero-temperature case,
if $\Gc_{PP}$ is sufficiently well-behaved in the chiral limit then
${\rm R}=0$, and spontaneous breaking of chiral symmetry by a finite
$\chicond$ leads to massless excitations in the spectrum. This is the
finite-temperature version of Goldstone's theorem
(see~\cite{Strocchi:2008gsa} and references therein). As shown below
in Section~\ref{sec:locpp}, localised near-zero modes can lead to a
nonzero ${\rm R}$, which can remove these Goldstone excitations from
the spectrum. Again, a finite anomalous remnant indicates explicit
breaking of chiral symmetry in the massless limit, so that the axial
current is not conserved and Goldstone's theorem at finite temperature
is evaded.

\section{Localised modes and the pseudoscalar correlator}
\label{sec:locpp}

I now show that a nonzero ${\rm R}$ is generally found in the presence
of a finite density of localised near-zero
modes~\cite{Giordano:2020twm}. Since UV divergences play a very
limited role, the argument can be carried out safely (and more simply)
in the continuum.  One starts from the bare pseudoscalar correlator
$\la \Pc_B^a(x)\Pc_B^b(0)\ra$ at temperature $T$ in a finite spatial
volume $V$ and for finite (bare) mass $m_B$,\footnote{The
  zero-temperature case is obtained by setting the calculation in a
  finite four-volume $V_4$, replacing $T/V\to 1/V_4$ in the formulas
  below, and eventually taking the limit $V_4\to\infty$.}  written in
terms of a double sum over Dirac modes,
\begin{equation}
  \label{eq:pp1}
          \la \Pc_B^a(x)\Pc_B^b(0)\ra
  =-\f{\delta^{ab}}{2} \left\la\sum_{n,n'}
    \f{\Oc^{\gamma_5}_{n'n}(x)\Oc^{\gamma_5}_{nn'}(0)}{(i\lambda_n
    +m_B)(i\lambda_{n'} +m_B)}\right\ra  \equiv
-\delta^{ab}  \Pi_B(x)\,. 
\end{equation}
Here $\slashed{D}\psi_n = i\lambda_n\psi_n$, with $\psi_n$ obeying
antiperiodic (resp.\ periodic) temporal (resp.\ spatial) boundary
conditions and normalised to 1,
$\Oc^\Gamma_{nn'}(x) \equiv \sum_{c,A,B}\psi_{n\,A,
  c}(x)^*\Gamma_{AB}\psi_{n'\,B, c}(x)$, and a UV cutoff on
$\lambda_{n,n'}$ is understood to be in place.  After renormalisation
of the mass, $m_B=Z_m m$, and of $\Pi_B$,
$\Pi(x) = Z_m^2[\Pi_B(x)-{\rm CT}(x)]$, including the removal of the
divergent contact terms ${\rm CT}$, one can take the thermodynamic and
chiral limit (in this order) to find the following expression for the
coefficient of the $1/m$ divergence of $\Pi(x)$,
\begin{align}
  \label{eq:pp3}
  \lim_{m\to 0} 2m\Pi(x) &=
                           2\lim_{m\to 0}\int_{0}^{\f{\mu}{m}} dz \left(\f{C^1(mz;m;x)}{z^2+1}
                           + \f{(1-z^2)C^{\gamma_5}(mz;m;x)}{(z^2+1)^2}
                           \right)\,,\\[1em]
    \label{eq:pp3_bis}
  C^{\Gamma}(\lambda;m;x) 
                         &  \equiv  \left\la {\textstyle\sum_n'}
                           \delta(\lambda-\lambda_{n}^R)
                           \Oc^{\Gamma}_{nn}(x)\Oc^{\Gamma}_{nn}(0)\right\ra\,,
\end{align}
where $\lambda^R_n = Z_m^{-1}\lambda_n$,
$\sum_n'=\sum_{\lambda_{n}^R\neq 0}$, and $\mu$ is a fixed but
arbitrary mass scale, which will eventually play no role. Exact zero
modes have been dropped since they are negligible in the thermodynamic
limit.  Modes outside of a neighbourhood of $\lambda=0$ also become
negligible in the chiral limit, leading in particular to the absence
of divergent contact terms.

The quantity in Eq.~\eqref{eq:pp3} can be nonvanishing only if
$C^{\Gamma}$ survives the thermodynamic limit, and here the localisation
properties of the eigenmodes play a crucial role. In fact, using
Schwarz inequality and translation invariance one can bound the
eigenmode correlators entering $C^\Gamma$ as follows,
\begin{equation}
  \label{eq:pp4}
  \begin{aligned}
    |\la \Oc^{\Gamma}_{nn}(x)\Oc^{\Gamma}_{nn}(0)\ra| &\le \la \Vert
    \psi_n(x) \Vert^2 \Vert \psi_n(0)\Vert^2\ra \le \f{1}{2}\la \Vert
    \psi_n(x) \Vert^4 + \Vert \psi_n(0)\Vert^4\ra \\ & = \f{T}{V}
    \left\la \int_T d^4x\, \Vert \psi_n(x) \Vert^4 \right\ra =
    \f{T}{V} \la {\rm IPR}_n\ra \,.
  \end{aligned}
\end{equation}
Making the dependence of $C^\Gamma$ on $V$ explicit by writing
$C_V^\Gamma$, one then finds
\begin{equation}
  \label{eq:pp5}
  |C_V^\Gamma(\lambda;m;x) |
  \le \f{T}{V}
  \left\la {\textstyle\sum_n'} 
    \delta(\lambda-\lambda^R_n)\, {\rm IPR}_n\right\ra
  = \rho_V(\lambda) \,
  \overline{\rm IPR}(\lambda)\,,
\end{equation}
where
$\rho_V(\lambda) \equiv \f{T}{V}\la\sum_n'
\delta(\lambda-\lambda^R_n)\ra$ and
$\overline{\rm IPR}(\lambda)\equiv \f{T}{V}\la \sum_n'
\delta(\lambda-\lambda^R_n)\, {\rm IPR}_n\ra /{\rho_V(\lambda)}$ are
the spectral density at finite $V$ and the average IPR computed
locally in the spectrum, respectively. If modes near $\lambda$ are
supported in a region of size $\Oc(V^{\alpha(\lambda)})$, one has
$\overline{\rm IPR}(\lambda) \sim V^{-\alpha(\lambda)}$, and so
$C^\Gamma_V(\lambda;m;x) \to 0$ in the thermodynamic limit unless
$\alpha(\lambda)=0$, i.e., unless modes near $\lambda$ are localised.
In the thermodynamic limit,
$ C^\Gamma(\lambda;m;x) = \lim_{V\to\infty} C^\Gamma_V(\lambda;m;x)$
is then nonzero only in spectral regions where localised modes are
present.

The anomalous remnant ${\rm R}$ is now obtained by integrating
Eq.~\eqref{eq:pp3} over Euclidean spacetime. Assuming that localised
modes are present in the interval $[0,\lambda_c(m)]$, one obtains
\begin{equation}
  \label{eq:pp7}
  {\rm R}  =     - \int_T d^4x  \lim_{m\to 0} 2m \Pi(x)
  = -\pi\xi\,\rho_{\rm loc}(0)\,,   
\end{equation}
where $\rho_{\rm loc}(0)$ is the density of localised near-zero modes,
\begin{equation}
  \label{eq:pp8}
  \rho_{\rm loc}(0) \equiv \lim_{m\to 0} \lim_{\lambda\to 0} \lim_{V\to
    \infty} \f{T}{V}\sum_{n\in {\rm loc}} \la\delta(\lambda-\lambda^R_n)\ra\,,
\end{equation}
and
\begin{equation}
  \label{eq:pp9}
\xi \equiv \lim_{m\to 0}  \f{2}{\pi}\arctan   \f{\lambda_c(m)}{m}  
\end{equation}
is a function of the renormalisation-group invariant ratio
$\f{\lambda_c(m)}{m} $ in the chiral limit. In obtaining
Eq.~\eqref{eq:pp7} one exploits the localised nature of the modes to
exchange the order of integration, chiral limit, and thermodynamic limit, as
well as the orthonormality of Dirac modes. As anticipated, ${\rm R} $
is proportional to the density of localised near-zero modes. The
quantity $\xi\in[0,1]$ depends on how the mobility edge scales in the
chiral limit: $\xi =0$ if it vanishes faster than $m$, $0<\xi<1$ if it
vanishes like $m$, and $\xi=1$ if it vanishes more slowly than $m$,
including not vanishing at all. The arbitrary scale $\mu$ does not
appear in the final expression, as expected.

\section{Localised modes and Goldstone excitations}
\label{sec:locG}

Using Eq.~\eqref{eq:pp7} and the Banks-Casher relation
$\chicond = -\pi\rho(0)$~\cite{Banks:1979yr}, where $\rho(0)$ is the
density of near-zero modes (localised or otherwise) in the chiral
limit obtained from $\rho_V(\lambda)$ taking limits as in
Eq.~\eqref{eq:pp8}, one finds for the singular part of the spectral
function in the chiral limit~\cite{Giordano:2020twm}
\begin{equation}
  \label{eq:gmodes1}
  \begin{aligned}
    \lim_{\vec{p}\to 0}\lim_{m\to 0}
    \varrho^{\Ac\Pc}(\omega,\vec{p})|_{\rm singular} &= -2\pi [
    \chicond - {\rm R} ]\delta(\omega) = 2\pi^2 \rho(0) \left( 1 - \xi
      \f{\rho_{\rm loc}(0)}{\rho(0)} \right)\delta(\omega) \,.
  \end{aligned}
\end{equation}
One can now determine the fate of the Goldstone excitations.  Since
localised and delocalised modes usually do not coexist, one has
$\f{\rho_{\rm loc}(0)}{\rho(0)}=1$ or $0$ depending on whether
near-zero modes are localised or delocalised. There are four possible
scenarios.
\begin{enumerate}
  \setcounter{enumi}{-1}
\item Near-zero modes are delocalised: Goldstone excitations are
  present as long as $\rho(0)\neq 0$. This is the standard scenario
  predicted by Goldstone's theorem.
\item Near-zero modes are localised and $\xi=0$: Goldstone excitations
  are present as long as $\rho(0)\neq 0$, i.e., localisation of
  near-zero modes has no effect on the Goldstone excitations, and the
  same standard scenario is found.
\item Near-zero modes are localised and $0<\xi< 1$: Goldstone
  excitations are present if $\rho(0)=\rho_{\rm loc}(0)\neq 0$,
  although the coefficient of the Dirac delta is reduced compared to
  scenarios 0 and 1.  This is qualitatively the same as the standard
  scenario, but differs from it quantitatively.
\item Near-zero modes are localised and $\xi= 1$: Goldstone
  excitations are absent even if $\rho(0)=\rho_{\rm loc}(0)\neq 0$.
\end{enumerate}

\section{Conclusions}
\label{sec:concl}

I have shown how the pseudoscalar-pseudoscalar correlator generally
develops a $1/m$ divergence in the chiral limit in the presence of a
finite density of localised near-zero modes. This divergence leads to
a finite anomalous remnant that modifies the usual form of the axial
nonsinglet Ward-Takahashi identity in the chiral limit, signaling that
chiral symmetry is broken explicitly even in this limit. This
indicates non-conservation of the axial-vector current, and so the
inapplicability of Goldstone's theorem, both at zero and at finite
temperature. Depending on the detailed behaviour of the mobility edge
$\lambda_c$ as a function of $m$, one can either recover the standard
scenario with massless excitations, possibly up to a change in the
coefficient of the singular term in the spectral function, or have
Goldstone excitations removed from the spectrum.

So far, the presence of localised near-zero modes in the chiral limit
has not been demonstrated explicitly in any model, although there are
indications that it could be a feature of the chiral limit of QCD and
of SU(3) gauge theory with $N_f=2$ flavours of adjoint fermions. It
would certainly be interesting to find a model with this property,
especially if it realised a non-standard scenario for Goldstone modes
(i.e., cases 2 and 3 above).  It would also be interesting to work out
the possible signatures in the finite-mass theory originating from the
realisation of a non-standard scenario in the chiral limit.

\vspace{-0.5em}
\begin{acknowledgments}
  This work was partially supported by the NKFIH grant KKP-126769.
\end{acknowledgments}


\vspace{-0.5em}

\begin{thebibliography}{10}

\bibitem{GarciaGarcia:2006gr}
A.M.~Garc\'ia-Garc\'ia and J.C.~Osborn, 
  \href{https://doi.org/10.1103/PhysRevD.75.034503}{\emph{Phys. Rev. D}
  {\bfseries 75} (2007) 034503}
  [\href{https://arxiv.org/abs/hep-lat/0611019}{{\ttfamily hep-lat/0611019}}].

\bibitem{Kovacs:2012zq}
T.G.~Kov{\'a}cs and F.~Pittler, 
  \href{https://doi.org/10.1103/PhysRevD.86.114515}{\emph{Phys. Rev. D}
  {\bfseries 86} (2012) 114515}
  [\href{https://arxiv.org/abs/1208.3475}{{\ttfamily 1208.3475}}].

\bibitem{Cossu:2016scb}
G.~Cossu and S.~Hashimoto, 
  \href{https://doi.org/10.1007/JHEP06(2016)056}{\emph{{J. High Energy Phys.}}
  {\bfseries 06} (2016) 056}
  [\href{https://arxiv.org/abs/1604.00768}{{\ttfamily 1604.00768}}].

\bibitem{Holicki:2018sms}
L.~Holicki, E.-M.~Ilgenfritz and L.~von Smekal, 
  \href{https://doi.org/10.22323/1.334.0180}{\emph{PoS} {\bfseries LATTICE2018}
  (2018) 180} [\href{https://arxiv.org/abs/1810.01130}{{\ttfamily
  1810.01130}}].

\bibitem{Kovacs:2009zj}
T.G.~Kov\'acs, 
  \href{https://doi.org/10.1103/PhysRevLett.104.031601}{\emph{Phys. Rev. Lett.}
  {\bfseries 104} (2010) 031601}
  [\href{https://arxiv.org/abs/0906.5373}{{\ttfamily 0906.5373}}].

\bibitem{Kovacs:2010wx}
T.G.~Kov\'acs and F.~Pittler, 
  \href{https://doi.org/10.1103/PhysRevLett.105.192001}{\emph{Phys.
  Rev. Lett.} {\bfseries 105} (2010) 192001}
  [\href{https://arxiv.org/abs/1006.1205}{{\ttfamily 1006.1205}}].

\bibitem{Giordano:2016nuu}
M.~Giordano, S.D.~Katz, T.G.~Kov\'acs and F.~Pittler, 
  \href{https://doi.org/10.1007/JHEP02(2017)055}{\emph{{J. High Energy Phys.}}
  {\bfseries 02} (2017) 055}
  [\href{https://arxiv.org/abs/1611.03284}{{\ttfamily 1611.03284}}].

\bibitem{Kovacs:2017uiz}
T.G.~Kov\'acs and R.\'A.~Vig, 
\href{https://doi.org/10.1103/PhysRevD.97.014502}{\emph{Phys. Rev.
  D} {\bfseries 97} (2018) 014502}
  [\href{https://arxiv.org/abs/1706.03562}{{\ttfamily 1706.03562}}].

\bibitem{Giordano:2019pvc}
M.~Giordano, 
  \href{https://doi.org/10.1007/JHEP05(2019)204}{\emph{{J. High Energy Phys.}}
  {\bfseries 05} (2019) 204}
  [\href{https://arxiv.org/abs/1903.04983}{{\ttfamily 1903.04983}}].

\bibitem{Vig:2020pgq}
R.\'A.~Vig and T.G.~Kov{\'a}cs, 
  \href{https://doi.org/10.1103/PhysRevD.101.094511}{\emph{Phys. Rev. D}
  {\bfseries 101} (2020) 094511}
  [\href{https://arxiv.org/abs/2001.06872}{{\ttfamily 2001.06872}}].

\bibitem{Bonati:2020lal}
C.~Bonati, M.~Cardinali, M.~D'Elia, M.~Giordano and F.~Mazziotti,
  \href{https://doi.org/10.1103/PhysRevD.103.034506}{\emph{Phys. Rev. D}
  {\bfseries 103} (2021) 034506}
  [\href{https://arxiv.org/abs/2012.13246}{{\ttfamily 2012.13246}}].

\bibitem{Baranka:2021san}
G.~Baranka and M.~Giordano, 
  \href{https://doi.org/10.1103/PhysRevD.104.054513}{\emph{Phys. Rev. D}
  {\bfseries 104} (2021) 054513}
  [\href{https://arxiv.org/abs/2104.03779}{{\ttfamily 2104.03779}}].

\bibitem{Cardinali:2021fpu}
M.~Cardinali, M.~D'Elia, F.~Garosi and M.~Giordano,
  \href{https://arxiv.org/abs/2110.10029}{{\ttfamily 2110.10029}}.

\bibitem{Giordano:2021qav}
M.~Giordano and T.G.~Kov{\'a}cs,
  \href{https://doi.org/10.3390/universe7060194}{\emph{Universe} {\bfseries 7}
  (2021) 194} [\href{https://arxiv.org/abs/2104.14388}{{\ttfamily
  2104.14388}}].

\bibitem{lee1985disordered}
P.A.~Lee and T.V.~Ramakrishnan, 
  \href{https://doi.org/10.1103/RevModPhys.57.287}{\emph{Rev. Mod. Phys.}
  {\bfseries 57} (1985) 287}.

\bibitem{Giordano:2013taa}
M.~Giordano, T.G.~Kov\'acs and F.~Pittler, 
  \href{https://doi.org/10.1103/PhysRevLett.112.102002}{\emph{Phys. Rev. Lett.}
  {\bfseries 112} (2014) 102002}
  [\href{https://arxiv.org/abs/1312.1179}{{\ttfamily 1312.1179}}].

\bibitem{Nishigaki:2013uya}
S.M.~Nishigaki, M.~Giordano, T.G.~Kov\'acs and F.~Pittler, 
  \href{https://doi.org/10.22323/1.187.0018}{\emph{PoS} {\bfseries LATTICE2013}
  (2014) 018} [\href{https://arxiv.org/abs/1312.3286}{{\ttfamily 1312.3286}}].
  
\bibitem{Ujfalusi:2015nha}
L.~Ujfalusi, M.~Giordano, F.~Pittler, T.G.~Kov\'acs and I.~Varga,
  \href{https://doi.org/10.1103/PhysRevD.92.094513}{\emph{Phys. Rev. D}
  {\bfseries 92} (2015) 094513}
  [\href{https://arxiv.org/abs/1507.02162}{{\ttfamily 1507.02162}}].

\bibitem{Evers:2008zz}
F.~Evers and A.D.~Mirlin, 
  \href{https://doi.org/10.1103/RevModPhys.80.1355}{\emph{Rev. Mod. Phys.}
  {\bfseries 80} (2008) 1355}
  [\href{https://arxiv.org/abs/0707.4378}{{\ttfamily 0707.4378}}].
  
\bibitem{Bruckmann:2011cc}
F.~Bruckmann, T.G.~Kov\'acs and S.~Schierenberg, 
  \href{https://doi.org/10.1103/PhysRevD.84.034505}{\emph{Phys. Rev. D}
  {\bfseries 84} (2011) 034505} 
  [\href{https://arxiv.org/abs/1105.5336}{{\ttfamily 1105.5336}}].

\bibitem{Giordano:2015vla}
M.~Giordano, T.G.~Kov\'acs and F.~Pittler, 
  \href{https://doi.org/10.1007/JHEP04(2015)112}{\emph{{J. High Energy Phys.}}
  {\bfseries 04} (2015) 112}
  [\href{https://arxiv.org/abs/1502.02532}{{\ttfamily 1502.02532}}].

\bibitem{Giordano:2016cjs}
M.~Giordano, T.G.~Kov\'acs and F.~Pittler, 
  \href{https://doi.org/10.1007/JHEP06(2016)007}{\emph{{J. High Energy Phys.}}
  {\bfseries 06} (2016) 007}
  [\href{https://arxiv.org/abs/1603.09548}{{\ttfamily 1603.09548}}].

\bibitem{Dick:2015twa}
V.~Dick, F.~Karsch, E.~Laermann, S.~Mukherjee and S.~Sharma, 
  \href{https://doi.org/10.1103/PhysRevD.91.094504}{\emph{Phys. Rev. D}
  {\bfseries 91} (2015) 094504}
  [\href{https://arxiv.org/abs/1502.06190}{{\ttfamily 1502.06190}}].

\bibitem{Ding:2020xlj}
H.-T.~Ding, S.-T.~Li, S.~Mukherjee, A.~Tomiya, X.-D.~Wang and Y.~Zhang,
  \href{https://doi.org/10.1103/PhysRevLett.126.082001}{\emph{Phys. Rev.
  Lett.} {\bfseries 126} (2021) 082001}
  [\href{https://arxiv.org/abs/2010.14836}{{\ttfamily 2010.14836}}].
  
\bibitem{Kaczmarek:2021ser}
O.~Kaczmarek, L.~Mazur and S.~Sharma, 
  \href{https://arxiv.org/abs/2102.06136}{{\ttfamily 2102.06136}}.

\bibitem{Karsch:1998qj}
F.~Karsch and M.~L{\"u}tgemeier, 
  \href{https://doi.org/10.1016/S0550-3213(99)00129-7}{\emph{Nucl. Phys. B}
  {\bfseries 550} (1999) 449}
  [\href{https://arxiv.org/abs/hep-lat/9812023}{{\ttfamily hep-lat/9812023}}].

\bibitem{Engels:2005te}
J.~Engels, S.~Holtmann and T.~Schulze, 
  \href{https://doi.org/10.1016/j.nuclphysb.2005.06.029}{\emph{Nucl. Phys. B}
  {\bfseries 724} (2005) 357}
  [\href{https://arxiv.org/abs/hep-lat/0505008}{{\ttfamily hep-lat/0505008}}].

\bibitem{Banks:1979yr}
T.~Banks and A.~Casher, 
  \href{https://doi.org/10.1016/0550-3213(80)90255-2}{\emph{Nucl.
  Phys. B} {\bfseries 169} (1980) 103}.

\bibitem{Goldstone:1962es}
J.~Goldstone, A.~Salam and S.~Weinberg, 
  \href{https://doi.org/10.1103/PhysRev.127.965}{\emph{Phys. Rev.} {\bfseries
  127} (1962) 965}.

\bibitem{McKane:1980fs}
A.J.~McKane and M.~Stone, 
  \href{https://doi.org/10.1016/0003-4916(81)90182-2}{\emph{Ann.
  Phys. (N.Y.)} {\bfseries 131} (1981) 36}.

\bibitem{Golterman:2003qe}
M.~Golterman and Y.~Shamir, 
  \href{https://doi.org/10.1103/PhysRevD.68.074501}{\emph{Phys. Rev. D}
  {\bfseries 68} (2003) 074501}
  [\href{https://arxiv.org/abs/hep-lat/0306002}{{\ttfamily hep-lat/0306002}}].

\bibitem{Meyer:2011gj}
H.B.~Meyer, 
  \href{https://doi.org/10.1140/epja/i2011-11086-3}{\emph{Eur. Phys. J. A}
  {\bfseries 47} (2011) 86} [\href{https://arxiv.org/abs/1104.3708}{{\ttfamily
  1104.3708}}].

\bibitem{Burnier:2017bod}
Y.~Burnier, H.-T.~Ding, O.~Kaczmarek, A.-L.~Kruse, M.~Laine, H.~Ohno et~al.,
  \href{https://doi.org/10.1007/JHEP11(2017)206}{\emph{{J. High Energy Phys.}}
  {\bfseries 11} (2017) 206}
  [\href{https://arxiv.org/abs/1709.07612}{{\ttfamily 1709.07612}}].

\bibitem{Giordano:2020twm}
M.~Giordano, 
  \href{https://doi.org/10.1088/1751-8121/ac1c3a}{\emph{J. Phys. A} {\bfseries
  54} (2021) 37LT01} [\href{https://arxiv.org/abs/2009.00486}{{\ttfamily
  2009.00486}}].

\bibitem{Strocchi:2008gsa}
F.~Strocchi, \emph{{Symmetry Breaking}}, vol.~732 of \emph{Lect. Notes Phys.},
  Springer, Berlin (2008).

\end{thebibliography}
\end{document}